%% file: dynamo3.tex
\documentstyle{europhys}
\input euromacr

\begin{document}

\euro{}{}{}{}
\Date{}
\shorttitle{P. GIULIANI \etal A NOTE ON SHELL MODELS ETC.}

\title{ A Note on Shell Models for MHD Turbulence}
\author{P. Giuliani \And V. 
    Carbone
     }
\institute{
Dipartimento di Fisica, Universit\'a della Calabria, 
87036 Rende (Cs), Italy}

\rec{}{}

\input epsf

\pacs{
\Pacs{47}{65+a}{Magnetohydrodynamic}
\Pacs{47}{27.GS}{Isotropic Turbulence}
       }
\maketitle

\begin{abstract}

We investigate the time evolution of two different (GOY-like) shell models 
which have been recently proposed to describe the gross features of MHD 
turbulence. We see that, even if they are formally of the same type sharing 
with MHD equations quadratic couplings and conserved quantities, differences 
exist which are related to conserved quantities. 

\end{abstract}

One of the striking features of turbulence in fluid flows is the existence of 
an energy cascade from large to small scales, which is usually described in the 
framework of the Richardson's picture \cite{F}. Some efforts to describe the cascade 
through toy models or phenomenological considerations have been made in the 
last years \cite{BJPV}, aimed to 
understand chaotic behaviour, energy spectra, intermittency, etc. Among 
others, Shell Models can be built up by dividing the wavevectors space $k$ into 
logarithmically spaced shells $\lambda^n\leq k/k_0 \leq \lambda^{n+1}$ 
($\lambda$ is usually taken equal to 2), each shell being characterized
 by a discrete wavevector 
$k_n = \lambda^n k_0$ ($n = 0,1,...,N$), and by one single 
dynamical variable whose evolution is representative of the dynamics of fields 
belonging to the shell. The equations for these variables can be written by 
assuming quadratic nonlinear couplings between shells (as the original 
equations), and some of the coupling coefficients can be fixed by invoking the 
conservation of invariants in the dissipationless and force-free case. Up 
to now the so called GOY-model \cite{G,JPV} has been considered as the 
most representative to reproduce the turbulent features in fluid flows. 

Shell Models for MHD turbulence have been introduced in the last years 
\cite{GLPG}, and in particular the properties of GOY--like models have been 
investigated by Biskamp \cite{B} and more recently by
Frick and Sokoloff \cite{FS} (herein after referred to as B model and FS model
respectively). Both models can be obtained as
particular cases of the following set of equations ($v_n$ and $b_n$ represent 
respectively the velocity and the magnetic field in dimensionless units)
\begin{eqnarray}
{du_n \over dt} = -\nu k_{n}^{2}u_{n}-\nu^{\prime} k_{n}^{-2}u_{n}+ik_{n}\Bigl\{
(u_{n+1}u_{n+2}-b_{n+1}b_{n+2}) \nonumber \\
- \frac{\delta}{\lambda}(u_{n-1}u_{n+1}-b_{n-1}b_{n+1})
- \frac{1-\delta}{\lambda^2}(u_{n-2}u_{n-1}-b_{n-2}b_{n-1})
\Bigr\}^{\ast}+f_{n}
\label{nonlinearev}
\end{eqnarray}
\begin{eqnarray}
{db_n \over dt} = -\eta k_{n}^{2}b_{n}+ik_n\Bigl\{
(1-\delta-\delta_m)(u_{n+1}b_{n+2}-b_{n+1}u_{n+2})\nonumber \\
+\frac{\delta_m}{\lambda}(u_{n-1}b_{n+1}-b_{n-1}u_{n+1})+
\frac{1-\delta_m}{\lambda ^{2}}(u_{n-2}b_{n-1}-b_{n-2}u_{n-1})
\Bigr\}^{\ast}
+g_n
\label{nonlineareb}
\end{eqnarray}
or, in terms of the complex 
Els\"asser variables $Z_n^{\pm}(t) = v_n(t) \pm b_n(t)$, particularly
useful in some solar--wind applications,
\begin{equation}
\label{nonlineare}
{dZ_n^{\pm} \over dt}=-\nu^{+}k_{n}^{2}Z_n^{\pm}
-\nu^{-}k_{n}^{2}Z_n^{\mp}
-\frac{\nu^{\prime}}{2}k_n^{-2}Z_n^{+}
-\frac{\nu^{\prime}}{2}k_n^{-2}Z_n^{-}
+ik_{n}T_{n}^{\pm *}+f_{n}^{\pm}
\end{equation}
where
\begin{eqnarray}
T_n^{\pm}             & = & \left\{  
{{\delta +\delta_{m}}\over 2}
Z_{n+1}^{\pm}Z_{n+2}^{\mp}+{{2-\delta -\delta_{m}}\over 2}
Z_{n+1}^{\mp}Z_{n+2}^{\pm} \right. \nonumber \\
                      &   &
\mbox{}+{{\delta_{m}-\delta}\over {2\lambda}}
Z_{n+1}^{\pm}Z_{n-1}^{\mp}-{{\delta +\delta_m}\over {2\lambda}}Z_{n+1}
^{\mp}Z_{n-1}^{\pm} \nonumber \\
                      &   &
\left. \mbox{}-{{\delta_{m}-\delta}\over{2\lambda^{2}}}
Z_{n-1}^{\pm}Z_{n-2}^{\mp}-
{{2-\delta -\delta_{m}}\over {2\lambda^{2}}}Z_{n-1}^{\mp}Z_{n-2}^{\pm}
\right\}
\end{eqnarray}
Here $\nu^{\pm}=(\nu\pm\eta)/2$,
being $\nu$ the kinematic viscosity and $\eta$ the resistivity,
$-\nu^{\prime} k_{n}^{-2}u_{n}$, eq. (\ref{nonlinearev}), is a drag term
specific to 2D cases (see below),
$f_n^{\pm}=(f_n\pm g_n)/2$ are external driving forces,
$\delta$ and $\delta_m$ are real
coupling coefficients to be determined.
In the inviscid unforced limit, equations (\ref{nonlineare}) conserve both
pseudoenergies
$
E^{\pm}(t) = (1/4)\sum_n \left|Z_n^{\pm}(t)\right|^2
$
for any value of $ \delta $ and $ \delta_{m} $ 
(the sum is extended to all the shells), which corresponds to the conservation 
of both the total energy 
$E = E^{+}+E^{-}=(1/2)\sum_n (\left|v_n(t)\right|^2+
\left|b_n(t)\right|^2)$
and the cross-helicity 
$h_C =  E^{+}-E^{-}=\sum_n Re(v_n b_n^*)$. The values of
$\delta$ and $\delta_{m}$ are fixed by imposing the conservation of 
another quantity which is the magnetic helicity in 3D or, in 2D, the mean
square magnetic potential \cite{B2}. In analogy with the fluid case \cite{DM}
we can define a generalized quantity as  
$ H_B^{(\alpha)}(t)=\sum_n (\mbox{sign}(\delta-1))^n
{|b_n(t)|^{2}}/{k_n^{\alpha}}
$
whose conservation implies $\delta = {1-\lambda^{-\alpha}}$,
$\delta_m =\lambda^{-\alpha}/(1+\lambda^{-\alpha})$ for
$\delta <1$, $0 <\delta_m<1$ and
$\delta   = {1+\lambda^{-\alpha}} $,
$\delta_m = -\lambda^{-\alpha}/(1-\lambda^{-\alpha})$ for
$\delta >1$, $\delta_m<0$, $\delta_m>1$. Thus two classes of
MHD GOY models can be defined with respect to the values of
$\delta$: 3D--like models for $\delta<1$, where $H_B^{(\alpha)}$ is not positive
definite; 2D--like models where $\delta>1$ and $H_B^{(\alpha)}$ is positive 
definite. This situation strongly resembles what happens in the hydrodynamic
case where 2D--like $(\delta>1)$ and 3D--like $(\delta<1)$ models are
conventionally distinguished with
respect to a second generalized conserved quantity
$H_K^{(\alpha)}(t)=
\sum_n {(\mbox{sign}(\delta-1))^n}k_n^{\alpha} \left|v_n(t)\right|^2.
$ Here the 3D and 2D cases are recovered for $\alpha=1,\,2$ where the ideal
invariants are identified respectively with kinetic helicity and enstrophy. It
should be noted that, although the hydrodynamic invariants are not conserved
in the magnetic case, the equations which link $\alpha$ and $\delta$ are
exactly the same for hydrodynamic and MHD models. Thus, once fixed $\alpha$ and $\delta$,
it is a simple matter to find out
which GOY model the MHD GOY one reduces to when $b_n=0$. 
To summarize we have that (with $\lambda=2$) the   
FS model for the 3D case is recovered for $\alpha=1$, $\delta=1/2$,
$\delta_m=1/3$ and reduces to the usual 3D GOY model for $b_n=0$.
The B 3D model is
actually a 2D--like model. It is obtained for $\alpha=1$, $\delta=3/2$,
$\delta_m=-1$ and reduces to a 2D--like GOY model that conserves a quantity
which has the same dimensions as kinetic helicity but is positive definite.
The 2D FS and B models coincide, they are recovered for $\alpha=2$,
$\delta=5/4$, $\delta_m=-1/3$
and reduce to the usual 2D GOY model for $b_n=0$.\\
Let us now briefly review the main results found in \cite{FS}.
The authors investigate
the problem of 
the magnetic field generation in a free-decaying turbulence, 
thus showing that: $1)$ in the $3D$ case magnetic energy grows and
reaches a value comparable with the kinetic one, in a way that the magnetic 
field growth
is unbounded in the kinematic case; $2)$ in the $2D$ case magnetic energy
slowly decays in the nonlinear as well as in the kinematic case. These results
have been interpreted as a 3D ``turbulent dynamo effect" \cite{FS} and seem to 
be in agreement with well-known results by which dynamo effect is not
possible in two dimensions \cite{Z}. Morover the authors investigate 
the spectral properties of the model in the stationary forced case,  
and they find that Kolmogorov spectral properties are 
only established when the cross-correlation is smaller than the energy of the
system. In this paper we are going to compare the results
obtained from the $3D$ B model and the $3D$ FS model as far as the 
``dynamo--like effect" is concerned.
\begin{figure}
\hbox{
\hfil\vbox{\epsfxsize=7 cm\epsffile{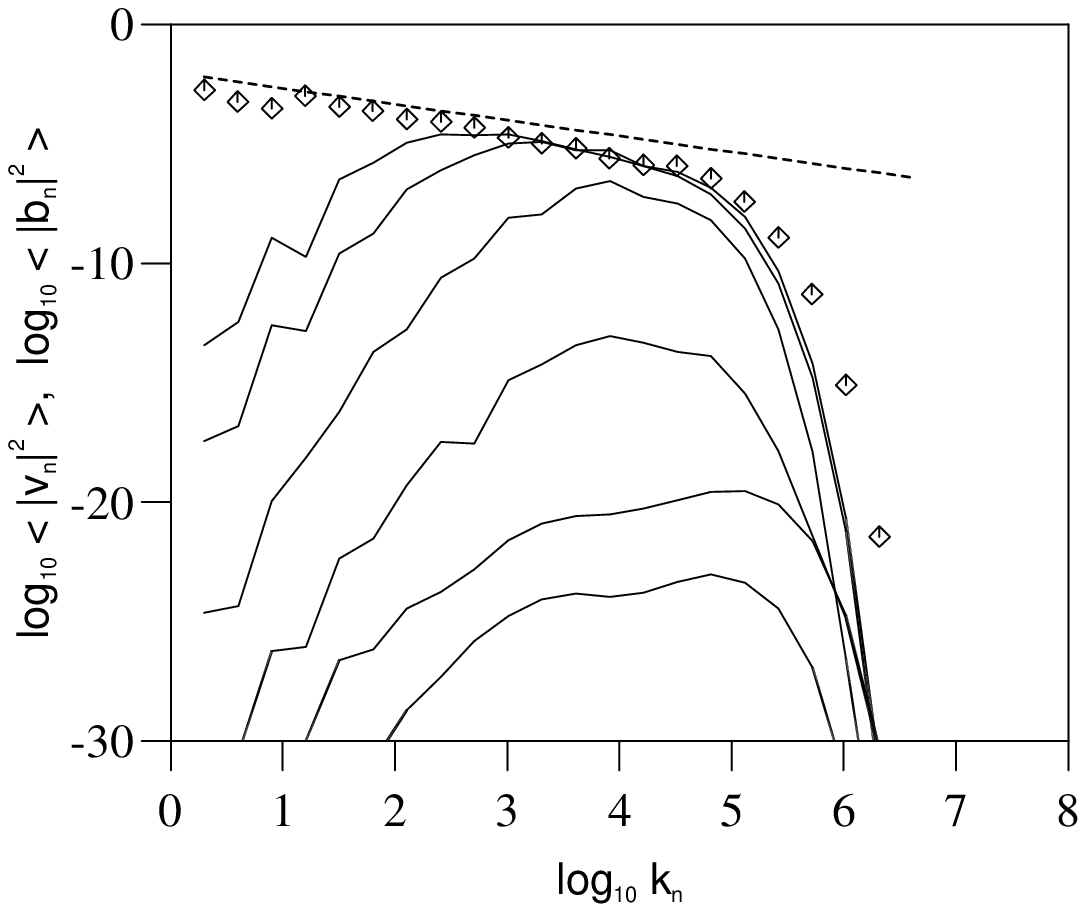}
\hbox{Fig. \ref{fseps}}}\hfil\hfil
\vbox{\epsfxsize=7 cm\epsffile{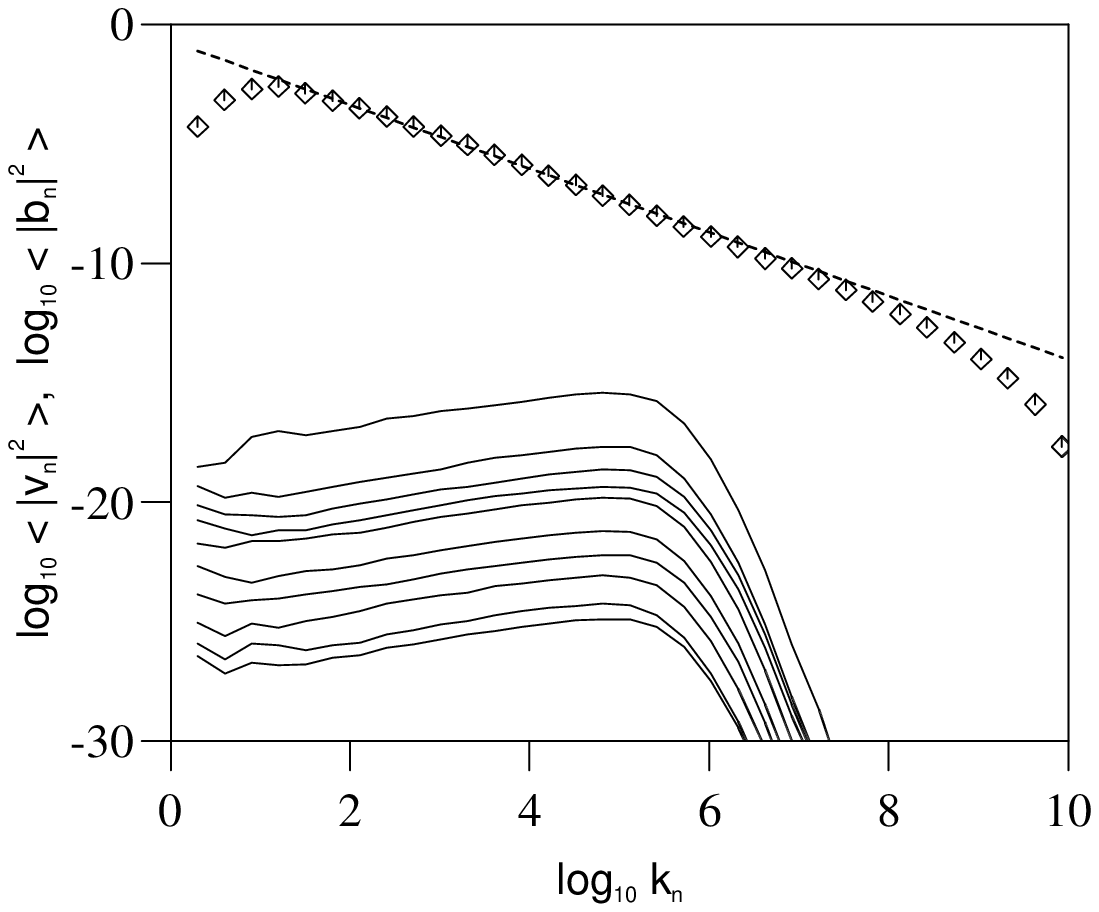}
\hbox{Fig. \ref{biskeps}}}\hfil
}
\caption{FS model: $\log_{10} \langle\, |v_n|^{2} \,\rangle$ (diamonds)
and $\log_{10}\langle\, |b_n|^{2}\,\rangle$ (lines) versus
$\log_{10}k_n$. 
The averages of $|b_n|^{2}$ are made over intervals of 3 large scale turnover 
times. Time proceeds upwards. The kinetic
spectrum is averaged over 30 large scale turnover times. The straight line has
slope $-2/3$. N=24, $\nu=\eta=10^{-8}$, $\nu^{\prime}=0$.\label{fseps}}
\caption{B model: $\log_{10} \langle\,|v_n|^{2}\,\rangle$ (diamonds)
and $\log_{10}\langle\,|b_n|^{2}\,\rangle$ (lines) versus 
$\log_{10}k_n$.
Averages are made over intervals of 100 large scale turnover times.
Time proceeds downwards. The kinetic
spectrum is only shown for the last interval. The straight line has
slope $-4/3$, see text for explanation. N=33, 
$\nu=10^{-16},\;\eta=0.5\cdot 10^{-9}$, $\nu^{\prime}=1$.\label{biskeps}}
\end{figure}
We started from a well developed turbulent
velocity field and injected a seed of magnetic field looking at the growth
of the magnetic spectra. System is forced on the shell $n=4$ ($k_0=1$), 
setting $f_4^{+}=f_4^{-}=(1+i)\,10^{-3}$, which corresponds to
only inject kinetic energy at large scales. Method of integration was a
modified fourth order Runge-Kutta scheme, with a time step $10^{-4}$. 
In fig. \ref{fseps} we report 
$\log_{10}\langle\,|b_n|^2\,\rangle$ and $\log_{10}\langle\,|v_n|^2\,\rangle$
versus $\log_{10}k_n$ for the FS model. Angular brackets $\langle\,\,\,\rangle$
stand for
time averages. It can be seen that the magnetic energy 
grows rapidly in time and forms a spectrum where the amplitude of the various 
modes is, at small scale, of the same order as the kinetic energy spectrum.
Besides the spectral index is very close
to $k^{-2/3}$.
For a comparison we integrated the 3D B model
and it can be seen (fig. \ref{biskeps}) that a magnetic spectrum is
formed, but it slowly decays
in time.
Notice that, because of the smallness of $b_n$, its back-reaction on the
velocity field is negligible, thus the kinematic part of the model
evolves independently from the magnetic one.
Then the scaling
$|v_n|^2\sim k_n^{-4/3}$ follows, in agreement 
with \cite{DM} where
a cascade of generalized-enstrophy is predicted for 2D--like hydrodynamic
GOY models 
when $\alpha<2$. 
\begin{figure}
\hfil\vbox{\epsfxsize=7 cm\epsffile{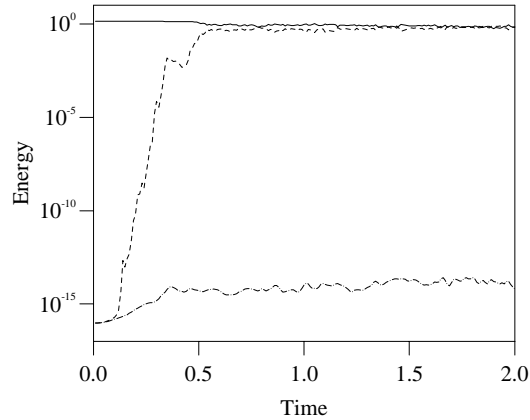}
}\hfil
\caption{Ideal case: kinetic energy (continuous line) and
magnetic energy (dashed line) versus time for the FS model; magnetic energy
(dot-dashed line) versus time for the B model.}
\label{eidealeps}
\end{figure}
The question now
arises whether it is correct the interpretation of the growth of the magnetic
field in the FS model as the corresponding dynamo
effect expected in the real 3D MHD.  First of all
it should be noted that in the kinematic case an analogy with the
vorticity equation predicts the following relations between velocity and
magnetic energy spectra \cite{B2}:
$|v_n|^{2}\sim\,k^{-a},\,|b_n|^{2}\sim\,k^{2-a}$,
so that
if $a=2/3$ it follows a magnetic energy spectrum growing with $k$.
The kinematic case corresponds to the first stage of growth of our simulation
where this
behaviour is sometimes visible, at least qualitatively. Note however that
the averages are made on very small time intervals because of the rapid growth
of the magnetic energy. A similar, much more pronounced behaviour is found for
the B model as
well. As regards the final state of evolution for the FS model, it should be  
reminded that
equipartition of magnetic and kinetic spectra at small scales is generally
attributed to
the Alfv\'en effect, that is the propagation in opposite directions
of turbulent eddies $Z_n^{+}$, $Z_n^{-}$ along the largest scale magnetic field
(see for example \cite{B2}). Note however that,
as already remarked in \cite{B} for
another kind of shell models, equipartition is reached
even if the Alfv\'enic terms are not included.
The reader is referred to \cite{B} for a
thorough discussion concerning the inclusion of Alfv\'enic terms in shell
models.

Actually we want to stress that: $1)$ The sign of the third ideal invariant 
seems to play a crucial role as far as the growth of small magnetic fields is 
concerned; $2)$ The stationary final state of
evolution for the FS model does {\it not} represent a Kolmogorov--like 
turbulence. In fact let us consider the ideal evolution of the model 
$dZ_n^{\pm}/dt = i k_n T_n^{\pm\ast}$. We can build up the phase space $S$ of 
dimension $D = 4N$, by 
using the Els\"asser variables as axis, so that a point in $S$ represents the 
system at a given time. A careful analysis of (\ref{nonlineare}) shows 
that there exist some subspaces $I \subset S$ of dimension $D=2N$
which remain invariant under 
the time evolution \cite{CV}. More formally, let $y(0)=(v_n,b_n)$ be a set of 
initial conditions such that $y(0) \in I$, $I$ is time invariant if the 
flow $T^t$, representing the time evolution operator in $S$, leaves 
$I$ invariant, that is $T^{t}[y(0)] = y(t) \in I$.
The kinetic subspace $K \subset S$, defined by $y(0) = (v_n,0)$ is
obviously the usual fluid GOY model. 
Further subspaces are the Alfv\'enic subspaces $A^{\pm}$ defined by 
$y(0)=(v_n,\pm v_n)$, say $Z_n^+ \not = 0$ and $Z_n^-=0$ (or vice versa). Each
initial condition in these subspaces is actually a fixed point
of the system. We studied the properties of stability of $K$ and $A^{\pm}$.
Following \cite{CV}, 
let us define for each $I$ the orthogonal complement $P$, namely
$S = I \oplus P$. Let us then decompose the solution as
$y(t)=(y_{int}(t),y_{ext}(t))$ where the subscripts refer to the $I$ and
$P$ subspaces respectively. Finally we can define the energies
$E_{int}=\|y_{int}\|^2$ and $E_{ext}=\|y_{ext}\|^2$. Note that the distance
of a point $y=(y_{int},y_{ext})$ from the subspace $I$ is
$d=\min\limits_{\hat{y} \in I}\|y-\hat{y}\|=\|y_{ext}\|$.
Then $E_{ext}$ represents the square of the distance 
of the solution from the invariant subspace. At time $t=0$, $E_{ext} = 
\epsilon E_{int}$ ($\epsilon \ll 1$) represents the energy of the 
perturbation. Since the total energy is constant in the ideal case, two
extreme situations can arise: 
$1)$ The external energy remains of the same order of its initial value, that 
is the solution is trapped near $I$ which is then a stable subspace; $2)$ The 
external energy assumes values of the same order as the internal energy, 
that is the solution is repelled away from the subspace which is then unstable. 
\begin{figure}
\hbox{
\hfil\vbox{\epsfxsize=7 cm\epsffile{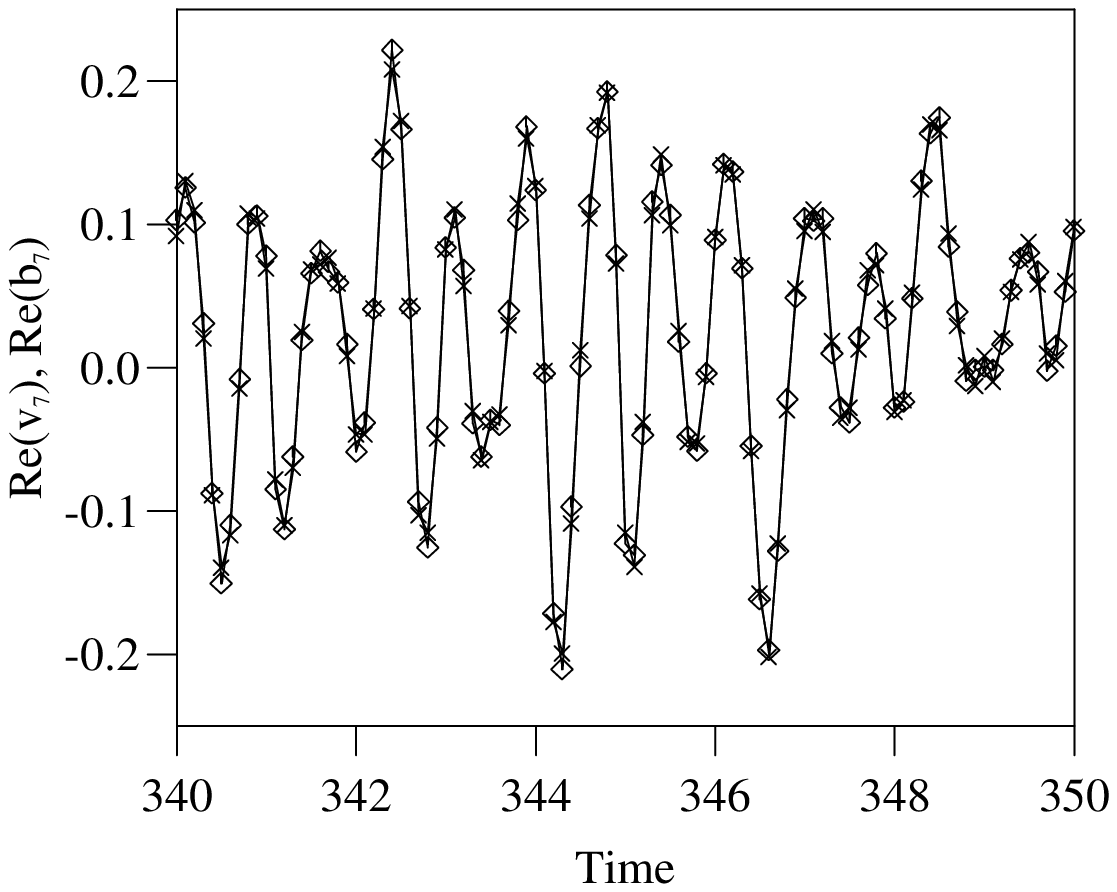}
\hbox{Fig. \ref{vbteps}}}\hfil\hfil
\vbox{\epsfxsize=7 cm\epsffile{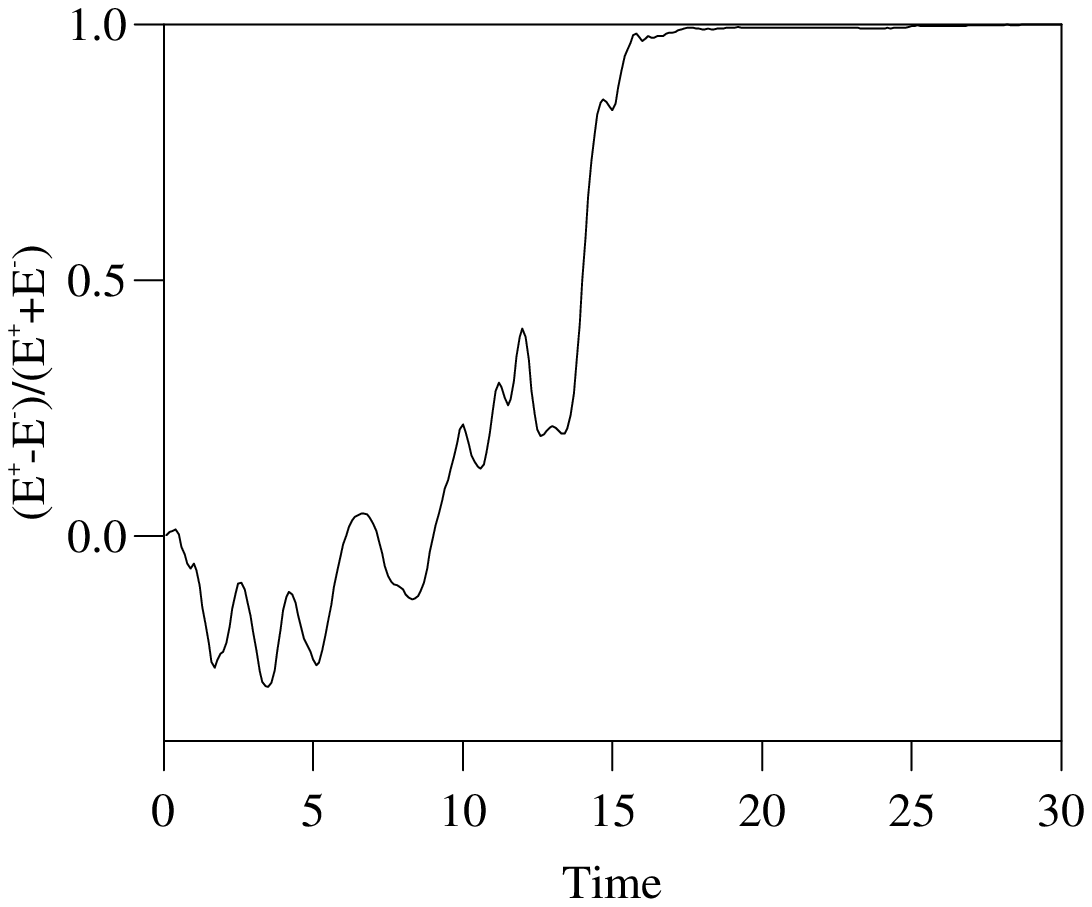}
\hbox{Fig. \ref{heeps}}}\hfil
}
\caption{FS model: $Re(v_{7})$ (diamonds) and 
$Re(b_{7})$ (crosses) versus time. N=19, $\nu=\eta=10^{-7}$, 
$\nu^{\prime}=0.$
\label{vbteps}}
\caption{FS model: $(E^{+}-E^{-})/(E^{+}+E^{-})$
versus time. Same parameters as in fig. 4.\label{heeps}}
\end{figure}
Since the external and internal energies 
for the Alfv\'enic subspaces are nothing but the pseudoenergies $E^+$ and 
$E^-$, which are ideal invariants, the Alfv\'enic subspaces are stable. As 
regards the kinetic subspace, $E_{int}$ and $E_{ext}$ represent respectively the 
kinetic and magnetic energies. Looking at the numerical solutions of the 
ideal model (fig. \ref{eidealeps}) we can see the difference in the stability 
properties between the B model and the FS model. In the first case the 
external energy remains approximately constant, while in the second case the 
system fills up immediately all the available phase space. This striking 
difference is entirely due to the nonlinear term, and in fact must be 
ascribed to the differences in sign of the third invariant.
The effect of the unstable subspace, which pushes away the solutions,
is what in ref. \cite{FS} is called ``turbulent dynamo effect".
When we introduce constant kinetic forcing and viscosity, the
stable subspaces become attractors, so that the B model is attracted by its
own kinetic subspace, while the FS model is attracted towards one of the
Alfv\'enic subspaces. The dynamics of the FS model is strongly dominated by 
this behavior. Long runs
show that the system evolves inevitably towards a ``dynamical alignment"
in which $v_n = \pm b_n$ as can be seen
in fig. \ref{vbteps} where we plot $Re({v_{7}})$ and $Re({b_{7}})$ versus time.
It is clearly visible how strongly velocity and magnetic field 
are correlated. 
In fig. \ref{heeps} we plot, for the FS model, the time evolution of the
reduced cross-helicity $(E^{+}-E^{-})/(E^{+}+E^{-})$, which is
a global measure of the dynamical alignment. It can be seen that
even from an initial value $h_C = 0$ it grows towards unity.
Due to the particular form of the
nonlinear interactions in MHD (\ref{nonlineare}), the nonlinear transfer towards
the smaller
scales tends to be stopped and $v_n$  and $b_n$ spectra become much
steeper than that reported in fig. \ref{fseps}. In this case the
turbulent statistical properties are not so clear as in a Kolmogorov--like
turbulence. The observed behaviour of the FS model might be avoided with
a careful choice of the external driving forces. A work in 
this perspective is actually in progress.
  
\stars
We are grateful to L. Biferale, G. Boffetta and A. Celani
for useful discussions, to 
P. Frick who gave us the preprint of his paper and to the referees
whose comments have improved the final version of the paper.

\end{document}

%% file: euromacr.tex
\def\etal{{\hbox{{\tenit\ et al.\/}\tenrm :\ }}}

\def\And{{\rm and\ }}

\def\stars{\bigskip\centerline{***}\medskip}

\newif\ifboo \boofalse

\def\Review#1{\boofalse{\it #1},}
\def\Name#1{{\sc #1},}
\def\Vol#1{\ifboo Vol. {\bf #1}\else{\bf #1}\fi}
\def\Year#1{\ifboo #1\else(#1)\fi}
\def\Book#1{\bootrue{\it #1},}
\def\Page#1{\ifboo {\rm p. #1}\else{\rm #1}\fi}